\documentclass{JINST}

\title{Evaluation of the optical cross talk level in the SiPMs adopted in ASTRI SST-2M Cherenkov Camera using EASIROC  front-end electronics }

\author{D. Impiombato$^a$, S. Giarrusso$^a$, T. Mineo$^a$, G. Agnetta$^a$, B. Biondo$^a$, O. Catalano$^a$, C. Gargano$^a$, G. La Rosa$^a$, F. Russo$^a$, G. Sottile$^a$, M. Belluso$^b$, S. Billotta$^b$, G. Bonanno$^b$, S. Garozzo$^b$, D. Marano$^b$, G. Romeo$^b$, on behalf of the ASTRI Collaboration \\ 
\llap{$^a$}INAF, Istituto di Astrofisica Spaziale e Fisica cosmica di Palermo,
via U. La Malfa 153, I-90146 Palermo, Italy\\
\llap{$^b$}INAF, Osservatorio Astrofisico di Catania, 
via S. Sofia 78, I-95123 Catania, Italy\\
E-mail: \email{Domenico.Impiombato@iasf-palermo.inaf.it,\\
 Salvatore.Giarrusso@iasf-palermo.inaf.it,\\
 Teresa.Mineo@iasf-palermo.inaf.it,\\
 Osvaldo.Catalano@iasf-palermo.inaf.it,\\
 Giovanni.LaRosa@iasf-palermo.inaf.it}}
 
\abstract{ASTRI (Astrofisica con Specchi a Tecnologia Replicante Italiana),
is a flagship project of the Italian Ministry of Education, 
University and Research  whose main goal 
  is the design and construction 
of an end-to-end prototype of the  Small Size of Telescopes 
of the Cherenkov Telescope Array. The prototype, named ASTRI SST-2M, 
will adopt a wide field dual mirror 
optical  system in a Schwarzschild-Couder configuration to explore 
the VHE  range of the electromagnetic spectrum. 
The camera at the focal plane is based on  
Silicon  Photo-Multipliers detectors which is an innovative 
solution for the detection  astronomical Cherenkov light.  
\\
This contribution reports some preliminary results on the evaluation 
of the optical cross talk level  among the SiPM pixels foreseen for 
the ASTRI SST-2M camera.}


\begin{document}

%%\thanks{Corresponding
%%author.}~ and 
\section{Introduction}
ASTRI (Astrofisica con Specchi a Tecnologia Replicante Italiana) \cite{Canestrari11},
is a flagship project of the Italian Ministry of Education, 
University and Research  led by the Italian National 
Institute of Astrophysics, INAF. 
Primary goal of the ASTRI project is the design 
and construction of an end-to-end prototype of the
of the small-size telescopes (SST)  
of the Cherenkov Telescope Array (CTA)\cite{Actis11}. 
The prototype, named ASTRI SST-2M, will adopt a wide field dual mirror 
optical  system in a Schwarzschild-Couder configuration to explore 
the VHE  range (1-100TeV) of the electromagnetic spectrum. 
The camera at the focal plane is based on Hamamatsu
 S11828-3344m\footnote{http://www.hamamatsu.com/sp/hpe/HamamatsuNews/HEN111.pdf} 
Silicon  Photo-Multipliers detectors which is an innovative 
solution for the detection of Cherenkov light  that requires 
high sensitivity in  the 300-700nm band and fast 
temporal response \cite{Catalano13}.
Each SiPM is a 4$\times$4 array of physical pixels that are 
grouped in 2$\times$2 logical pixels 
 of size of 0.17$^\circ$ in order  to match the optics angular resolution
 (see Figure 1).
 
 The SiPMs adopted for the ASTRI SST-2M camera will
be read by the front-end CITIROC (Cherenkov Imaging Telescope 
Integrated Read Out Chip) whose precursor 
EASIROC (Extended Analogue Silicon Photo-Multiplier Integrated Read
Out Chip) has been used
to perform the measurements in this paper.
 
EASIROC is equipped with 32-channels each with the capability of 
measuring charge from 0.3 to 2000 photoelectrons. 

To verify that the solutions 
adopted for the camera electronics and the choice of the detectors 
are compliant with the ASTRI SST-2M requirements,  a number of tests 
were carried out at the INAF laboratories in Palermo and in Catania 
\cite{Catalano13,ImpiombatoScineghe13,Impiombato13,Sottile13,Marano13,Marano14}.
In this paper we present some preliminary results on the evaluation 
of the optical cross talk level  in the SiPM pixel array foreseen for 
the ASTRI SST-2M  camera.

\begin{figure*}[!h]
\centering
\includegraphics[angle=0, width=13cm]{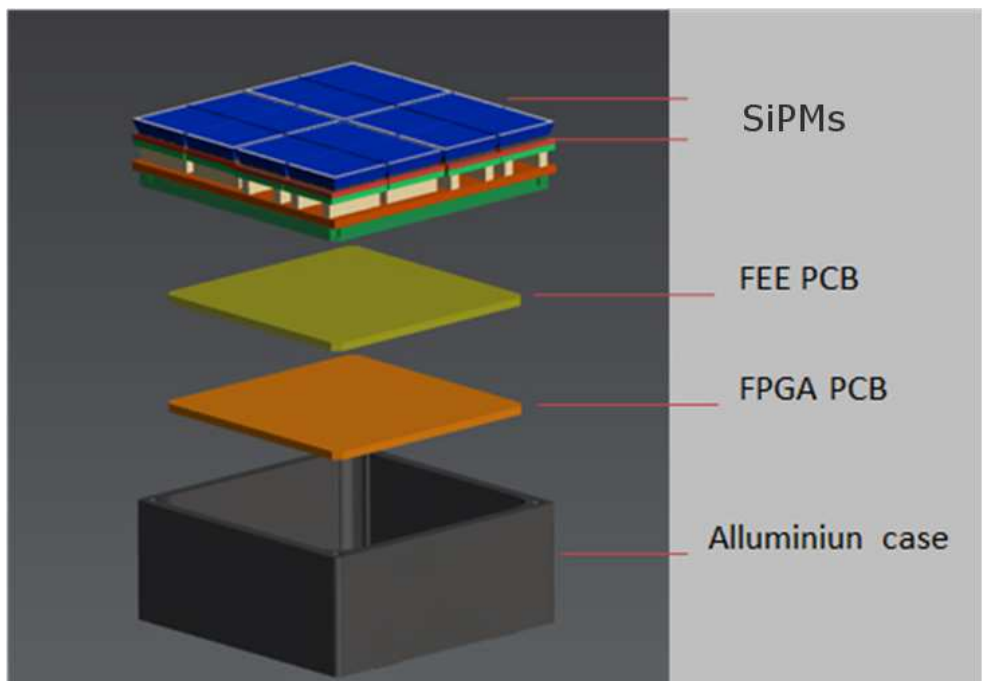}
\caption{Exploded view of the PDM mechanical module.}
\label{fig1}
\end{figure*}

\section{EASIROC Front-End}

 EASIROC, precursor of CITIROC,  was used to 
obtain a preliminary evaluation of the cross talk level
in the SiPM adopted for the ASTRI SST-2M camera.

EASIROC \cite{Callier12} is a 32 channel fully analogue front end 
ASIC(Application Specific Integrated Circuit) 
dedicated to readout SiPM detectors 
 specifically developed by the institute 
IN2P3-CNRS and the firm Omega Micro\footnote{http://omega.in2p3.fr} (France).   
An Evaluation Board designed and realized by
 Omega Micro allows 
to test the functional characteristics and performance of  the ASIC.
Two separate chains, high and low gain respectively, are 
implemented in the ASIC in order 
to measure charge from 0.3 photo-electron (pe) up to 2000pe.
 Each of the two chains is composed by an adjustable gain 
 preamplifier followed
by a tunable shaper and a track and hold circuit. 
A shaping time of 50 ns has been adopted for both 
the low and high gain chain for the pulse height
measurements. A third chain is implemented to 
generate a trigger using a fast shaper
(15ns) followed by a discriminator with  
adjustable threshold set by a 10-bit DAC
(Digital to Analog Converter) common to all 32 channels.

\begin{figure*}[!h]
\centering
\includegraphics[angle=0, width=15cm]{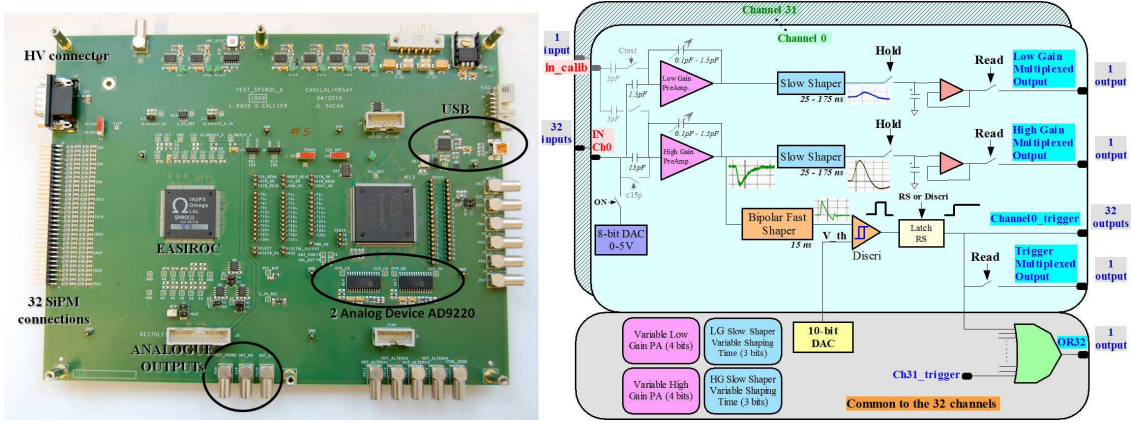}
\caption{ Architecture of the front-end EASIROC (Omega Micro courtesy)}
\label{fig2}
\end{figure*}

\newpage
\section{Experimental Setups}
In the measurements presented in this work, the physical 
SiPM pixel was directly connected to
 a channel of the ASIC; the over-voltage was set to 0.88V.
 
  A LabView based software developed by the 
 Omega Test group\footnote{ http://www.lal.in2p3.fr/}, 
 provided with the evaluation board, implements 
 all the required functions including ASIC configuration 
 set-up and data taking.

Measurements were performed at room temperature ($\sim$24$^\circ$C) 
without any temperature control.

\section{Cross Talk for physical SiPM pixel}

Two different methods were used to evaluate the SiPM
cross talk level: the first method is based on
a scan of the dark noise pulses at different trigger thresholds  
and the second one evaluates the cross talk from the pulse height 
distribution of the signal.

\subsection{Method 1: Cross Talk evaluation from the trigger chain}
This method is based on measurements of  the trigger rate as 
a function of the discriminator threshold in dark noise 
regime \cite{Eckert10}.
Data were accumulated  for 10s.  
The results are shown in the figure 3 where the characteristic 
staircase function is clearly evident: the count rate drops every 
time integer  multiples of 1 pe are reached.
The rate of the first plateau, that is relative to discriminator 
thresholds < 1pe,  gives the total dark noise rate, 
that in our measurements is  $\sim9.8\cdot10^{5}$ Hz. 
 Assuming a Poisson distribution, the probability to have two dark noise 
 coincident  events within a time window of 15ns is about $10^{-2}\%$ 
 (rate ~100 Hz), negligible with respect 
 to the dark rate measured.
According to  reference \cite{Eckert10}, the cross talk level is then 
evaluated  from the ratio 
$P_{c} = \nu_{1.5 pe}/\nu_{0.5 pe}$, where the approximated 
values of $\nu_{0.5 pe}$ and $\nu_{1.5 pe}$ are taken from 
the point marked in the Figure 3  with the red arrows 
that are representative of the average values in the two plateaus.
With these assumptions a cross talk  probability 
of  $\sim$ 24 \% is obtained. 

\begin{figure*}[!h]
\centering
\includegraphics[angle=0, width=12cm]{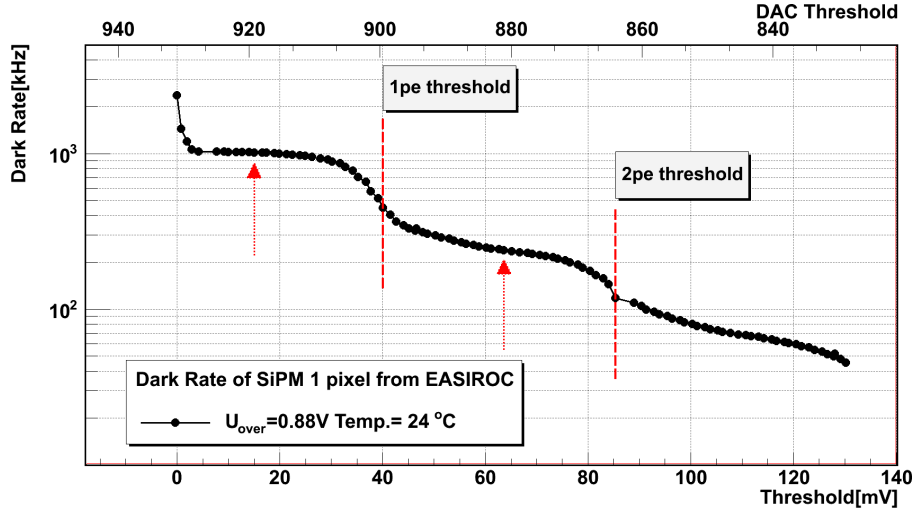}
\caption{Thermal noise rate of a 1 pixel SiPM  Hamamatsu S11828-3344m operated at an over-voltage of 0.88V as a function of 
the discriminator threshold.  The red arrows indicate the
threshold from where the cross level was computed (see text).
The dashed red lines mark the threshold level relative to 1pe and 2pe.}
\label{fig3}
\end{figure*}

\newpage
\subsection{Method 2: Cross talk from  pulse height distribution}

The dark count SiPM pulse height distribution was obtained using 
the High Gain electronics chain setting the shaping time to 
50ns and collecting a number of events of $\sim10^{5}$.
These measurements include also the effects of the afterpulses 
that are not observed in the method 1 because of the shorter 
integration time.
 The results are shown  in  Figure 4, where the peak relative 
 to the pedestal (0pe), 1pe , and 2pe are significantly resolved.
We first evaluated the afterpulses contribution in  the spectral 
interval between 1pe and 2pe. 
We fitted  with a Gaussian the 1pe peak and found that it is well 
modeled with a sigma $\sim$4.4 ADC (Analog to Digital Converter)  unit.
To fit the event distribution relative to 2pe we fixed the sigma 
at the value 6.2 ADC($\surd2\cdot4.4$), leaving free the other two parameters. 
 The complete fitting model, continuous red line, 
is presented in figure 4.
 The afterpulse contribution $P_{afterpulse}$ in the 50ns shaping time window
 is obtained from the difference between the  number of events detected 
 in the spectral region bounded by  the two red dashed lines 
 and  the integrated counts on the 2pe Gaussian fitted curve.
We find that  the afterpulse contribution is $P_{afterpulse} \cong14\%$. 
\\ 
The distribution shows also the presence of an excess of counts 
in the higher ADC channel  due to cross talk with 
multiple photo-electron($\geqslant3$).
 Assuming that the contribution of the
afterpulse is equal to the level measured between 1 and 2pe, 
we evaluated the total cross talk contribution from the 
following equation:

\begin{equation}
\label{fm1}
 P_{crosstalk}\,=\,\frac{(1-P_{afterpulse})\times\sum_{k=910}^{1000} count_{k}}{\sum_{k=840}^{1000} count_{k}}
\end{equation} 
\noindent
 where $P_{crosstalk}$ is the cross talk probability, 
$count_{k}$ is the number of events for each ADC unit.

We find a cross talk value $\sim22\%$.

\begin{figure*}[!h]
\centering
\includegraphics[angle=0, width=12cm]{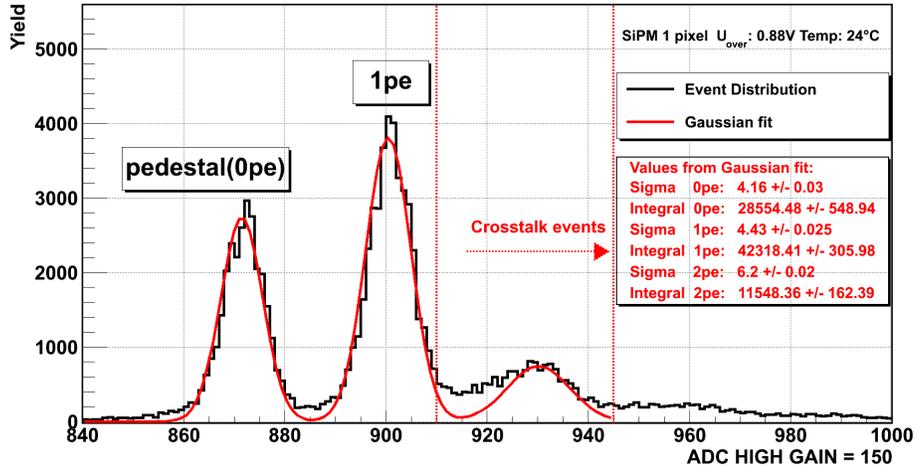}
\caption{ Sample charge histogram recorded for the uniformity scan 
measurement.  The red line shows the fitting model.}
\label{fig4}
\end{figure*}

\section{Summary and Conclusion}

 In this paper we focused our attention in developing
independent methods to evaluate the cross talk level.
The measurements we presented were performed at room temperature
without any temperature control among different SiPM pixel.
The two methods considered for the investigation give comparable 
cross talk levels($22\%-24\%$).
\\
The ASTRI SST-2M prototype will be tested on field in
Italy: the installation is foreseen in 2014 at the INAF
"M.G. Fracastoro"\cite{Maccarone13} observing station in Serra La Nave 
near Catania.
The camera will operate at the controlled 
temperature of  about 15$^\circ$C and more extensive tests
 and measurements are in progress to evaluate the 
 cross talk level in real operating condition.
Cross talk is a characteristic feature of the SiPM technology adopted.
The new generation of SiPMs show a cross talk level of a few percent.

\acknowledgments

The work presented in this paper was partially supported by the ASTRI,
"Flagship Project" financed by the Italian Ministry of Education, University,
and Research (MIUR) and  led by the Italian National Institute 
of Astrophysics (INAF). 
We also acknowledge partial support by the MIUR Bando PRIN 2009.
We are deeply grateful to S. Callier, C. De La Taille,
and L. Raux of the Omega Micro at Orsay and to
ASTRI collaborators for useful discussions and suggestions.

\end{document}